\documentstyle[aps,floats,epsfig]{revtex}
\begin{document}
\draft
\twocolumn[\hsize\textwidth\columnwidth\hsize\csname @twocolumnfalse\endcsname
\title{Band-theoretical prediction of magnetic anisotropy in uranium
monochalcogenides}
\author{Tatsuya Shishidou}
\address{Japan Science and Technology Corporation, Tokyo 102-0081, Japan}
\address{and Department of Quantum Matter, ADSM, Hiroshima University,
Higashihiroshima 739-8526, Japan}
\author{Tamio Oguchi}
\address{Department of Quantum Matter, ADSM, Hiroshima University,
Higashihiroshima 739-8526, Japan}
\date{\today}
\maketitle
\begin{abstract}
Magnetic anisotropy of uranium monochalcogenides, US, USe and UTe, is
studied by means of
fully-relativistic spin-polarized band structure calculations within the
local spin-density 
approximation. 
It is found that the size of the magnetic anisotropy is fairly large 
($\simeq$10 meV/unit formula), which is comparable with experiment. 
This strong anisotropy is discussed in view of a pseudo-gap formation,
of which crucial ingredients are the exchange splitting of U $5f$ states and their hybridization 
with chalcogen $p$ states ($f$--$p$ hybridization). 
An anomalous trend in the anisotropy is found in the series (US$\gg$USe$<$UTe) 
and interpreted in terms of competition between localization of the U $5f$
states and the $f$--$p$ hybridization. 
It is the spin-orbit interaction on the chalcogen $p$ states that plays an
essential role 
in enlarging the strength of the $f$--$p$ hybridization in UTe, 
leading to an anomalous systematic trend in the magnetic anisotropy. 
\end{abstract}
% insert suggested PACS numbers in braces on next line
]

Magnetic moments in solid originate in the spin and orbital components
of electrons. 
Since electronic states responsible for the magnetism are normally
localized in a particular atomic region, 
the moments can be regarded as site-selective quantities. 
Experimental techniques such as x-ray magnetic circular dichroism (XMCD)
combined with 
the so-called spin and orbital sum rules \cite{Thole,Carra} provide such
separable information of the spin and orbital magnetic 
moments of ferro- and ferri-magnets. 
X-ray magnetic scattering \cite{Blume} can also
give us similar information. 

Usually $5f$ electrons play major roles in magnetism of uranium compounds. 
Since the spin-orbit interaction (SOI) of the $5f$ electrons is relatively
large, 
the size of the $5f$ orbital moment is often expected to be greater than
the spin counterpart. 
Unlike the $4f$ orbitals in rare-earth element systems, 
the $5f$ states are more or less extended and may be possibly 
affected by their environmental effects, hybridization and crystal field. 
Thus, the magnetic moment in $5f$ systems must be strongly
material-dependent. Magnetic anisotropy is another fundamental quantity  
in magnetism, which is often even more important for applications. 
However, the magnetic anisotropy energy is usually very small to be 
evaluated from first principles 
and furthermore its microscopic origins have never been clearly 
understood yet \cite{Jansen}. 
It is, therefore, quite interesting to study such issues on the magnetism 
by using a state-of-the-art band-theoretical technique. 

Uranium monochalcogenides, US, USe and UTe, have NaCl-type cubic crystal
structure 
and show a ferromagnetic order at the Curie temperatures, 177K, 160K and
104K, respectively\cite{Handbook Actinide}. 
It is well known that the monochalcogenides show strong magnetic
anisotropy \cite{Lander}, 
where the [111] ([001]) direction is the easy (hard) axis. 
Interestingly, the saturation magnetic moment depends on the magnetization
axis \cite{Tillwick}. 
The largest moment along the easy axis, the smallest along the hard axis. 
The total magnetic moment per uranium atom increases from sulfide through
telluride with increasing lattice constant. 
As long as the sulfide is concerned, the $5f$ electrons are considered to
be itinerant from photoemission \cite{Reihl} and other
experiments \cite{Schoenes,Huang,Rudigier}. 

In the present study, mechanism of the magnetic anisotropy in the uranium
monochalcogenides is investigated by 
first-principles calculations and an anomalous trend in the size of the
anisotropy in the series is predicted. 
Our method is based on the local spin-density approximation (LSDA) to the
density functional theory. 
One-electron Kohn-Sham equations are solved self-consistently by using 
an iterative scheme of the
full-potential 
linear augmented plane wave method \cite{Soler} in a scalar-relativistic
fashion. 
We include SOI as the second variation every self-consistent-field step. 
The improved tetrahedron method proposed by Bl\"{o}chl \cite{Blochl} is used
for the Brillouin-zone integration. 
More details about our methods and calculated results are 
published elsewhere \cite{Shishidou2}.
% fig_band %%%%%%%%%%%%%%%%%%%%%%%%%%%%%%%%%%%%%%%%%%
\begin{figure}[htb]
\centerline{\epsfig{file=band.epsi,width=8cm}} 
%\centerline{\epsfxsize=8.0cm\epsfbox{band.epsi}}
\vspace{3mm}
\caption{Fully-relativistic spin-polarized band structure of ferromagnetic
US with the [111] 
magnetization. The Fermi energy is chosen as the origin.}
\label{fig_band}
\end{figure}
% fig_band %%%%%%%%%%%%%%%%%%%%%%%%%%%%%%%%%%%%%%%%%%

Figure \ref{fig_band} shows calculated fully-relativistic spin-polarized
band structure of ferromagnetic US with the [111] magnetization. 
Shallow core U $6p$ states form $j=1/2$ and $3/2$ bands split by large SOI.  
S $3s$ bands are located just above the U $6p_{3/2}$ bands and have
relatively large dispersion. 
S $3p$ bands are situated from 7 to 3 eV below the Fermi energy
($\varepsilon_{\rm F}$). 
Dispersive bands appearing just below $\varepsilon_{\rm F}$ are made mostly
of U $d$ states. 
Relatively narrow U $5f$ bands are pinned around $\varepsilon_{\rm F}$. 
It is found that the largest contribution to the magnetic moments comes
from U $5f$ states as expected. 
Calculated $5f$ moments of US are listed in Table \ref{tab1} for 
three magnetization axes, [001],  [110] and [111]. 
%TABLE1%%%%%%%%%%%%%%%%%%%%%%%%%%%%%%%%%%%%%%%%%%
\begin{table}[htdp]
\caption{Calculated $5f$ spin and orbital magnetic moments 
in Bohr magneton of US for different magnetization directions ($\hat{M}$). 
$\Delta E$ is a difference in the total energy in meV from the case of the
[111] magnetization.}
\begin{tabular}{ccccc}
$\hat{M}$ & spin & orbital & total & $\Delta E$ \\
\hline \\
$[001]$\tablenotemark[1] &1.61 & -2.13 & -0.52 & 14 \\
$[110]$\tablenotemark[1] &1.60 & -2.24 & -0.64 & 5 \\
$[111]$\tablenotemark[1] &1.60 & -2.33 & -0.73 & 0 \\
%LAPW & [111] &1.7 & -2.6 & -0.9 & -1.5 & \\
%ASW & [111] & 1.5 & -2.6 & -1.1 & -1.7 & \\
%LMTO & [111] & 2.1 & -3.2 & -1.1 & -1.5 & \\
%
$[111]$\tablenotemark[2]  & 1.3\tablenotemark[3] & -3.0\tablenotemark[3] &
-1.7 & \\
\end{tabular}
\tablenotetext[1]{Present work.}
\tablenotetext[2]{Neutron-scattering experiment (Ref.\ \onlinecite{Wedgwood}).}
\tablenotetext[3]{From analysis in Ref.\ \onlinecite{Severin}.}
\label{tab1}
\end{table}%
%TABLE1%%%%%%%%%%%%%%%%%%%%%%%%%%%%%%%%%%%%%%%%%%
% fig_jzdos %%%%%%%%%%%%%%%%%%%%%%%%%%%%%%%%%%%%%%%%%%
\begin{figure*}[bht]
\centerline{\epsfig{file=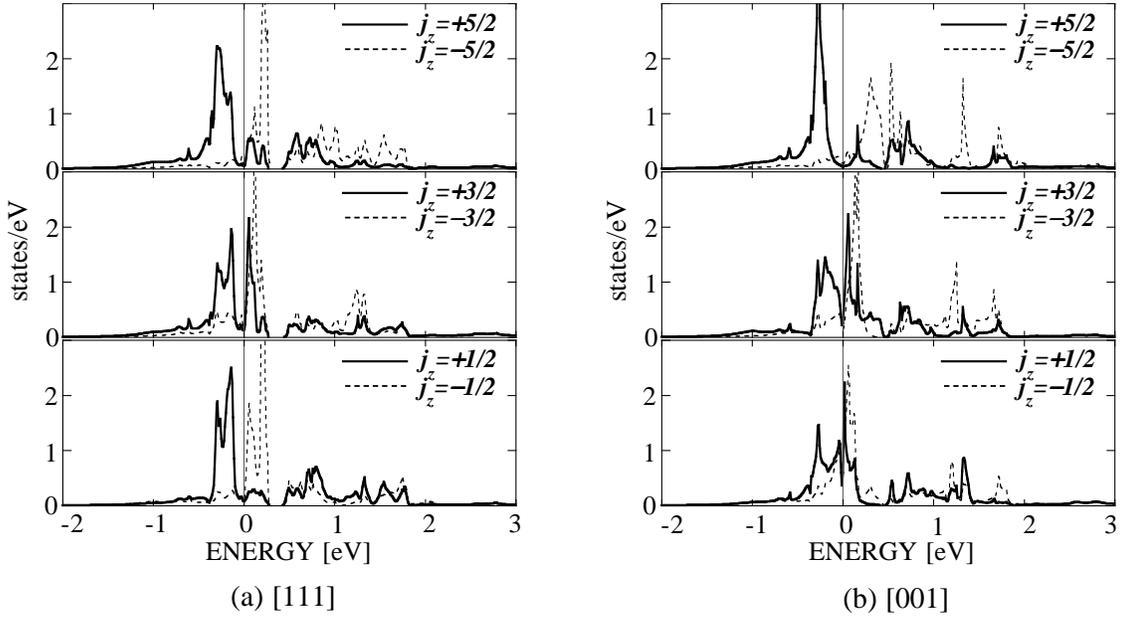,width=16cm}} 
\vspace{3mm}
\caption{Calculated $j_z$-projected $5f$ density of states ($j=5/2$) of
ferromagnetic US 
with the (a) [111] and (b) [001] magnetization directions. 
Quantization axis for specifying $j_z$ is taken as the same direction of
magnetization.}
\label{fig_jzdos}
\end{figure*}
% fig_jzdos %%%%%%%%%%%%%%%%%%%%%%%%%%%%%%%%%%%%%%%%%%
The spin moments hardly change with the magnetization axis 
while the orbital moments show large dependence of the axis. 
Consequently, the total moments depend strongly on the magnetization axis, 
being in qualitatively good agreement with experiment \cite{Tillwick}.  
%The results for the easy axis is consistent with the previously calculated ones. 
%The discrepancy may be due to the difference in the sphere radius assumed. 
Comparing with the experimental data, the spin moment is slightly overestimated 
while the orbital moment is underestimated. 
Origins of the discrepancy in the magnetic moments have been already 
discussed in the previous Hartree-Fock type study \cite{Shishidou}. 
The total-energy difference shows that the [111] direction is the lowest 
while [001] is the highest. 
The difference of the total energy between the easy and hard axes is nothing but 
the magnetic anisotropy energy. 
The size of the magnetic anisotropy is comparable with experimental 
data \cite{Lander} and is about $10^3$ times larger than that of 
$3d$ transition metals. 
Let us discuss mechanism of the magnetic anisotropy and its relation to the
magnetic moments.

Because of large SOI of $5f$ electrons, $j=5/2$ and $7/2$ states are 
roughly well separated 
and the occupied states are composed mostly of the $j=5/2$ states. 
The $j_z$-projected density of states (DOS) in the $j=5/2$ states is plotted in
Fig.~\ref{fig_jzdos}. 
One can easily note that a clear pseudo-gap is formed at $\varepsilon_{\rm F}$ 
in the [111] magnetization, 
while it becomes less apparent in the [001] direction. 
Relative stability of the [111] magnetization must come from the existence 
of the pseudo-gap. 

In the case of [111], the plus and minus components are well separated. 
This is the exchange splitting of the bands and the pseudo-gap is
considered to be a sort of the exchange gap. 
Only a small amount of mixing can be seen in the $j_{z} = \pm 3/2$ states.
On the other hand, the pseudo-gap is almost diminished in the case of [001] 
because of larger mixing in $j_{z}= \pm 3/2$ and $j_{z}= \pm 1/2$. 
By counting the number of occupied electrons in each partial DOS, 
one can get the occupation of each $j_z$ state. 
For [111] $\pm j_z$ state are well exchange-polarized 
but for [001] the occupations in the $j_{z}= \pm 1/2$ bands are almost
equal due to the mixing. 

\begin{figure*}[bht]
\centerline{\epsfig{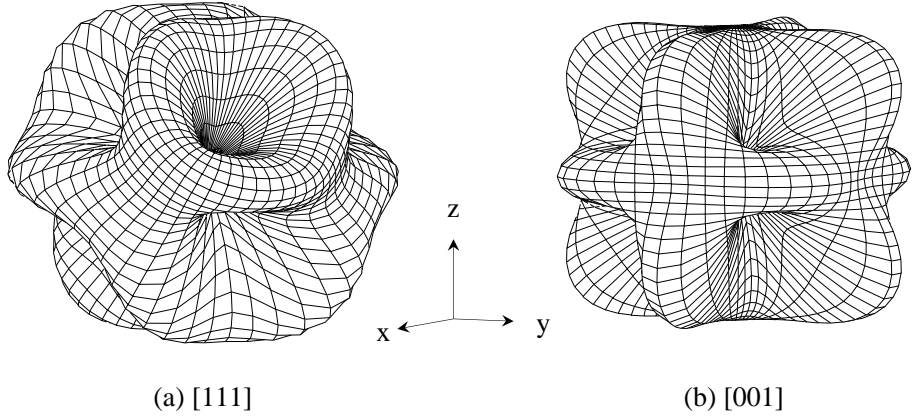}} 
\vspace{3mm}
\caption{Angular dependence of calculated charge density of the $5f$ electrons 
around the U atom with the (a) [111] and (b) [001] magnetization directions. 
The spherical component has been reduced to $20\%$ of its value 
to emphasize nonspherical parts.}
\label{fig_charge}
\end{figure*}

In order to get more intuitive insight into the occupations, 
a change-density plot of the $5f$ electrons around the U atom is 
depicted in Fig.~\ref{fig_charge}. 
% fig_charge %%%%%%%%%%%%%%%%%%%%%%%%%%%%%%%%%%%%%%%%%%
% fig_charge %%%%%%%%%%%%%%%%%%%%%%%%%%%%%%%%%%%%%%%%%%
In the [111] magnetization, the $5f$ electrons tend to point to 
the direction of neighboring-U atoms; 
hexagonal brim stretching out in the (111) U plane 
and triangular bell around [111] direction are clearly formed. 
As a result, the $5f$ electrons have less distribution in the neighboring-S
directions 
and they can reduce the hybridization with the S $p$ states and gain the 
exchange splitting. 
The small mixing found in $j_{z}=\pm 3/2$ can be understood by the fact
that the 
$j_{z}=\pm 3/2$ orbitals are extended to the S atoms 
with respect to the polar angle \cite{quantization}.  
In the [001] magnetization, on the other hand, 
the $5f$ electrons again try to point to the U-U bonds, 
twelve [110] directions. 
In the $xy$ plane gentle depression is formed in the nearest-neighboring-S
directions, 
because the $j_{z}=\pm 5/2$ states which spread in the $xy$ plane 
can prevent from mixing with the S $p$ states by using their azimuthal
degrees of freedom. 
However, no depression is seen in the [$00\pm 1$] directions. 
The $j_{z}=\pm 1/2$ states extending to the $z$ direction 
have no way to refuse the mixing with the S $p$ states 
because of less azimuthal degrees of freedom in its $m=0$ component. 
This hybridization of the $j_{z}=\pm 1/2$ states with the neighboring S $p$
orbits 
destroys the pseudo-gap in the corresponding partial band and 
makes the [001] magnetization unfavorable \cite{S_111}. 
This is the most important mechanism of the magnetic anisotropy 
found in the present study and leads to a very interesting variation 
in the series of the uranium monochalcogenide as we shall discuss below. 

\begin{figure}[htb]
\centerline{\epsfig{file=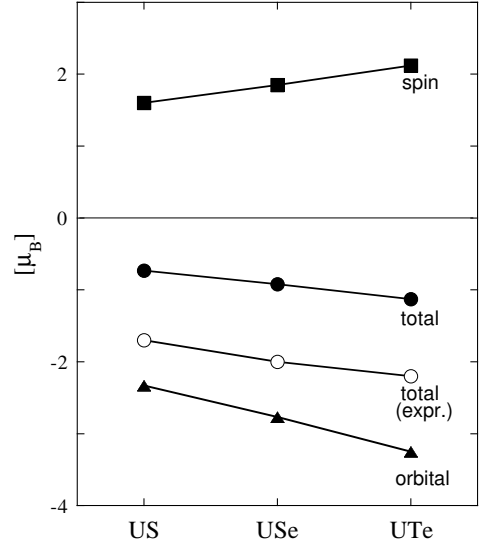,width=8cm}} 
\caption{Calculated spin and orbital magnetic moments in $\mu_{\rm B}$ of US, USe and UTe 
with the [111] magnetization. 
Filled rectangles, triangles, and circles represent 
the spin, orbital, and total $5f$ magnetic moments, respectively.
Open circles show experimental $5f$ moments \protect\cite{Wedgwood}.}
%(Ref.\ \onlinecite{Wedgwood}).}
\label{fig_moment}
\end{figure}

\begin{figure}[htb]
\centerline{\epsfig{file=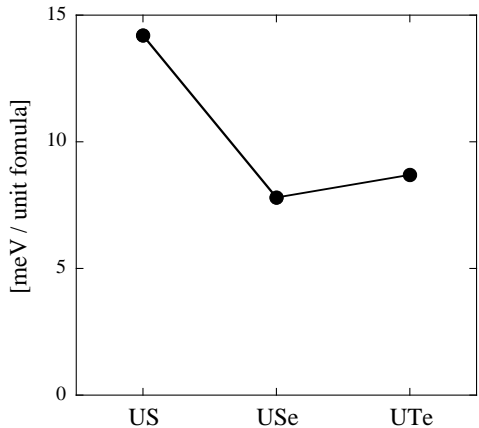,width=8cm}} 
\caption{Calculated magnetic anisotropy energy of US, USe and UTe, taken from the 
total-energy difference between the [111] and [001] magnetization directions.}
\label{fig_aniene}
\end{figure}

A variation in the calculated spin and orbital magnetic moments 
for USe and UTe as well as for US is shown in Fig.~\ref{fig_moment}. 
% fig_monent %%%%%%%%%%%%%%%%%%%%%%%%%%%%%%%%%%%%%%%%%%
% fig_moment %%%%%%%%%%%%%%%%%%%%%%%%%%%%%%%%%%%%%%%%%%
The spin and orbital moments increase as one goes from sulfide through
telluride. 
This increase of the moments can be interpreted by narrowing of the U $5f$ 
bands due to increase of the lattice constant. 
Based on such theoretical observation on the $5f$-electron nature, 
we can expect a simple trend in the magnetic anisotropy. 
From sulfide to telluride, the lattice constant increases and the $5f$ band
width is reduced. 
Accordingly the magnetic moments are enhanced, approaching to an atomic limit. 
Therefore, the magnetic anisotropy may show a {\em monotonic decrease\/} 
from sulfide through telluride with increasing free-atom nature. 
But the reality is not so simple and calculated magnetic anisotropy energy, 
plotted in Fig.~\ref{fig_aniene}, shows an anomalous behavior: {\em
decrease and increase\/} with 
a minimum appearing at USe. 
% fig_aniene %%%%%%%%%%%%%%%%%%%%%%%%%%%%%%%%%%%%%%%%%%
% fig_aniene %%%%%%%%%%%%%%%%%%%%%%%%%%%%%%%%%%%%%%%%%%
Calling that the strong magnetic anisotropy in the present compounds 
originates in the hybridization between the U $5f$ and chalcogen $p$ states 
($f$--$p$ hybridization), 
differences in the chalcogen $p$ states may be a clue to understand such an 
anomalous feature. 

Hybridization depends upon spatial extension of the relevant orbitals via
transfer integrals. 
Surprisingly, tails of the chalcogen $p$ orbitals are almost the same 
if we consider the nearest neighbor distance between the U and chalcogen atoms. 
The Te $5p_{3/2}$ orbital has the longest tail of the chalcogens 
but the difference is not so large. 
In addition to the orbital extension, the orbital energy is another factor
to determine the hybridization. 
Generally, the orbital-energy difference between the U $5f$ and chalcogen
$p$ states becomes smaller 
when going from sulfur through tellurium. 
Because of the large SOI, the Te $5p_{3/2}$ state raises up and its
orbital-energy difference from the 
U $5f$ decreases substantially. 
Therefore, the $f$--$p$ hybridization 
becomes very strong in the case of telluride. 
Basically, from sulfide through telluride, 
the U $5f$ electrons tend to be more localized due to enlargement of the
lattice constant. 
But the $f$--$p$ hybridization, which makes the [001] magnetization unfavorable, 
increases rapidly from selenide to telluride, resulting to large magnetic
anisotropy in UTe. 
It should be noted that experimental data about the magnetic anisotropy 
have not been available for USe and UTe \cite{Lander2}.  
Experimental efforts to examine our prediction are strongly desired. 

In conclusion, we have carried out fully-relativistic LSDA calculations for
uranium monochalcogenides, US, USe and UTe. 
The magnetic anisotropy can be well described by the LSDA calculations. 
The pseudo-gap formation stabilizes the [111] magnetization. 
We have emphasized important roles of the $j_{z}=\pm 1/2$ states, 
which make the [001] magnetization unfavorable significantly. 
Calculated magnetic moments show the monotonic increase in the series 
with increasing lattice constant. 
On the other hand, the magnetic anisotropy shows the anomalous systematic trend. 
This can be understood by considering the hybridization of 
the U $5f$ states with the chalcogen $p$, especially the $p_{3/2}$ components. 
Anyhow, SOI not only on U but also on the chalcogen sites 
is essential to realize such interesting nature in magnetism of the uranium 
chalcogenides. 

We thank Takeo Jo for invaluable discussion. 
This work was supported in part by Japan Science and Technology Corporation.

\begin{references}
\bibitem{quantization}Note that $j_z$ is defined by taking the quantization
axis 
as the same direction as the magnetization.
\bibitem{S_111}As for the [111] magnetization 
there also exist the S atoms in the magnetization direction, 
but they are the 3rd nearest neighbors and the hybridization effect is much 
weaker.
\end{references}
\end{document}